# Revisiting 70 years of lattice dynamics of BaTiO$_3$: Combined first principle and experimental investigation

Vivek Dwij[1], Binoy Krishna De[1], Gaurav Sharma[1], D. K. Shukla[1], M. K. Gupta[2], R. Mittal[2,3], Vasant Sathe[1,*]

[1]UGC-DAE-Consortium for Scientific Research, D. A. University Campus, Khandwa Road, Indore-452001, India

[2]Solid State Physics Division, Bhabha Atomic Research Center, Mumbai 400 085, India

[3]Homi Bhabha National Institute, Anushaktinagar, Mumbai 400094, India

*Corresponding author: vasant@csr.res.in

*Abstract*

BaTiO$_3$ is a classical ferroelectric studied for last one century for its ferroelectric properties. Lattice dynamics of BaTiO$_3$ is crucial as the utility of devices is governed by phonons. In this work, we show that traditional characterization of the polar phonon modes is ambiguous and often misinterpreted. By combining Raman, Neutron and X-ray diffraction, dielectric spectroscopic observations with first principle calculations, we have re-examined the character of the normal modes of phonons of BaTiO$_3$. We obtained Eigen displacements of vibrational modes through DFT calculations and reclassified the polar modes being Slater (Ti-O), Last (Ba-TiO$_3$) and Axe (BO$_6$) vibrations by correlating experimental and theoretical calculations. The study thus provides correct nomenclature of the polar modes along with the evidence of presence of short range polar distortions along (111) directions in all the phases shown by BaTiO$_3$. The Burn's temperature and absence of second order contributions have been witnessed in the temperature dependent Raman study.

# 1. Introduction

BaTiO$_3$ is a classical ferroelectric compound studied intensely for several decades. It has been currently researched for its usages in the field of piezoelectrics, ferroelectric memory (FeRAM) and photovoltaics etc. [1,2]. Recently, the control of the domain wall motion using polarized light has been demonstrated in the BaTiO$_3$ [3] using Raman spectroscopy. The control of domain walls using polarized light can eliminate the need of electrical circuitry and thus can lead to faster and efficient FeRAM devices. However, correct identification of vibrational modes of BaTiO$_3$ and the vibrational signatures of ferroelectricity are still contentious. It is important to understand and correctly identify the different vibrational features as they ascribe to the electromechanical tensor and thermodynamic properties of the BaTiO$_3$. Even after a large number of theoretical and experimental studies on Raman modes, vibrational features are described differently in existing literature. Presence of broad modes in the Raman spectrum had caused a controversy way back in the 1980s [4]. In the tetragonal phase of the BaTiO$_3$, the factor group theoretical calculations results in $\Gamma=3A_1+4E+B_1$ zone-center $\Gamma$-point optical phonon modes (the details are given in supplementary material Table S1 [5]) which further shows a splitting of polar modes in LO and TO braches in ferroelectric tetragonal phase. The degenerate E(TO) and non-degenerate A(TO) modes are polarized in x-y plane and along z axis contributing to dielectric constant $\varepsilon_a$ and $\varepsilon_c$ respectively. It is well known that there are 3(E+A$_1$) polar modes in the tetragonal phase of the BaTiO$_3$ which are named after Slater, Last and Axe [6,7,8] and are related to Ti-O, Ba-TiO$_3$ and bending type BO$_6$ vibrations. The Slater mode represents the vibrations of Ti in the Oxygen octahedra and becomes active in polar state due to shifting of Ti fromcentrosymmetric position, the Last mode represents the oscillation of A-cations against the BO$_6$–octahedra framework while the Axe mode arises due to the bending of O$_6$ octahedra. In the literature, the position of the polar modes and its vibrational characters are defined differently by different workers. For examples, Last [7] predicted the Ba-TiO$_3$ mode around 220 cm$^{-1}$ while Axe referred the mode lying around 180 cm$^{-1}$ as the Last mode. Perry and Hall [9] claimed that the mode lying at 270 cm$^{-1}$ as LO component of the low lying soft mode. DiDomenico *et al* [10] did not observed the A(TO) component of the soft mode, however, they predicted it to be around 180 cm$^{-1}$ using Lyddane-Sachs-Teller relation. Freire and Katiyar [11] calculated eigen displacements of BaTiO$_3$. They classified normal modes at 38 and180 cm$^{-1}$ as E/A(TO$_1$) modes, 180 and 270 cm$^{-1}$ as E/A(TO$_2$) modes, 305 and 520 cm$^{-1}$ as the E/A(TO$_3$) modes involving (Ti-O), (Ba-TiO$_3$) and (Ti-O$_6$) type vibrations, respectively. For iso-structural KNbO$_3$ [12,13], TO$_1$, TO$_2$ and TO$_3$ are associated with identical displacements in cubic phase.

The classification by Hermet and Raeliarijaona et al [14,15] differed from Freire and Katiyar and termed $E(TO_1)/A(TO_2)$ modes occurring at 38, 270 cm$^{-1}$ as (Ti-O) Slater type vibrations, $E(TO_2)/A(TO_1)$ modes at 180 cm$^{-1}$ as (Ba-TiO$_3$) Last type vibrations and $E(TO_4)/A(TO_3)$ at 490 and 520 as Axe (Ti-O$_6$) type vibrations. Ghosez and I. Ponomareva et al [16,17] defined $E(TO_1)$ mode as Slater type displacement while in the Zr doped BaTiO$_3$, Dobal et al [18] associated $A(TO_1)$ mode with the Ti atomic vibrations i.e. Slater type vibrations. On the contrary, Hlinka et al [19] used the notations Last, Slater and Axe for $A(TO_1)$, $A(TO_2)$ and $A(TO_3)$ modes lying at 180, 270, 520 cm$^{-1}$, respectively. Nuzhnyy et al [20] suggested that the lowest wave number modes for Zr and Ti sublattices should be Last and Slater type but the calculations of Wang et al [21] on Zr doped BaTiO$_3$ resulted into an intermediate type vibrations (i.e. superposition of Last and Slater type modes) for all the modes below 300 cm$^{-1}$. Buscaglia et al [22] (in Zr doped BaTiO$_3$) suggested that the $A(TO_2)$, $A(TO_3)$ are related to the polar Ti-O vibrations and defined the $A/E(LO_4)$ phonon modes being breathing and stretching type vibration of the BO$_6$ octahedra and correlated them with the ferroelectricity. Rubio-Marcos et al [3] attributed the ($A(TO_1)$, $E(TO_2)$, $E(LO_1)$, $A(LO_1)$) modes at 180 cm$^{-1}$ as the vibrations of Ti against the oxygen octahedral i.e. Slater type vibrations. They reckoned $A(TO_2)$ mode to be around 210-220 cm$^{-1}$ and used these modes for obtaining information about the domain wall motion. Poojitha et al [23] treated the $A(TO_2)$ mode as Last type displacements and B$_1$ mode as a signature of the ferroelectricity. While, Datta et al [24] used the notation of A-BO$_3$ (Last) type vibration for the soft $E(TO_1)$ phonon and related $A(TO_2)$ with Ti-O vibrations. In summary, the current understanding and characterization of the polar modes of the BaTiO$_3$ is highly ambiguous. Motivated by the lack of understanding, here we focus on realizing the correct characterization of the polar modes of BaTiO$_3$. Combining first principle calculations with experimental observations we show that $E(TO_1)$ and $A(TO_2)$ are Slater modes, $E(TO_2)$, $A(TO_1)$ are Last modes, $E(TO_4)$ and $A(TO_3)$ are Axe modes. The conventionally associated polar mode 305 cm$^{-1}$ has been examined and 3 distinct contributions are decoupled. Our interpretations of Raman modes are irrespective of domain state, microstructure, synthesis conditions etc.

## 2. Experimental Section

We have used a commercially obtained one side polished multi-domain single crystal BaTiO$_3$ (100) substrate (10×10×0.5 mm$^3$) for the experiments. The single crystal was cleaned using aceton before characterization. The Raman spectroscopic measurements were performed using Horiba JY HR-800 spectrometer equipped with 1800 g/mm grating and a CCD detector. We used a diode laser (473 nm) beam as an excitation source, which was focused onto ~1μm diameter spot in the backscattering geometry. We used an Olympus microscope to view the images of the surface of the sample and an LMplanFI 50× objective lens for focusing. The overall spectral resolution of the system is better than 1 cm$^{-1}$. The temperature dependent Raman measurements were performed by mounting the single crystal in a Linkam THMS600 stage for the low temperature measurements (80-340 K) and in a Linkam TS1000 stage for the high temperature measurements (300-1018 K), with temperature stability of ±0.1 K. Temperature was increased at 10K/min rate and the spectrum was collected after waiting for 2 min duration of reaching the temperature for ensuring thermal equilibrium between the stage and the sample. Polarized Raman spectra were recorded in both parallel and cross polarization configuration. Vacuum sputtered Au electrodes were applied on both sides of the polished single crystal for in-situ electric field dependent Raman and dielectric studies. In-situ electric field dependent Raman study of the Single crystal BaTiO$_3$ was carried out by manually varying electric field from 0- 2.62kV/cm. Dielectric studies were performed by utilizing HIOKI LCR meter (IM 3536) and CRYOCON temperature controller. Vacuum sputtered Au electrodes were applied on both sides of the polished single crystal. The ramp rate of 1K/min was kept for the temperature variations and data was collected for heating cycle in the temperature window of 80-490K and 300-750K separately under application of 1V.

Polycrystalline PbTiO$_3$, K$_{0.5}$Bi$_{0.5}$TiO$_3$ samples were prepared by solid state reaction method, using high purity (>99.9%) PbO, K$_2$CO$_3$, Bi$_2$O$_3$, TiO$_2$ as precursors weighted and mixed in stoichiometry and grinded in mortar using pestle. The grinded mixture of PbO+TiO$_2$ with 5% excess PbO was calcined at 600°C for 10 hours. After grinding calcined powder was pressed into pallet and sintered at 900°c under Lead rich environment of PbZrO$_3$. The grinded mixture of K$_2$CO$_3$+Bi$_2$O$_3$+TiO$_2$ with 5% excess Bi$_2$O$_3$ was calcined at 800°C for 10 hours. After grinding calcined powder was pressed into pallet and sintered at 1050°C for 5 hours under Bismuth rich environment. All of the pallets were prepared by mixing with poly vinyl alcohol (PVA) and then pressed into pellets under uniaxial pressure. X-ray diffraction was carried out using Cu K$_α$ radiation (Laboratory Rigaku diffractometer) with step size of 0.02°. For diffraction, the pallets were grinded and the grinded powder was heated above T$_c$ for 30 mins to remove any residual stresses due to grinding. Details of the Neutron diffraction and synthesis of polycrystalline BaTiO$_3$ are reported in the [25]. In the current work, the Raw data obtained from the same has been refined using Rietveld Algorithm [26] separately from the previous investigation and details have been given in supplementary [5].

# 3. Results

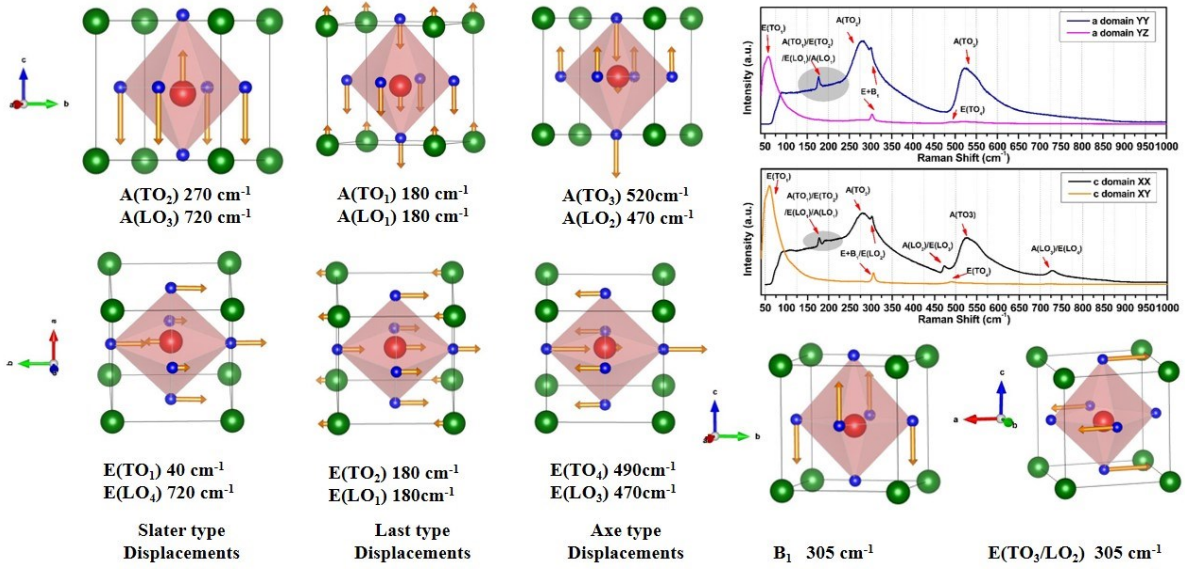

Fig. 1: Representative displacements of atoms in normal modes of BaTiO$_3$ obtained from the first principle calculations (drawn using VESTA software) [27] along with polarized Raman spectra collected on single crystal of BaTiO$_3$. The arrow on the atoms indicate the direction of the displacement and its magnitude is proportional to the displacement in the individual mode.

To confirm the origin of the atomic displacements in the phonon modes of BaTiO$_3$, we carried out first principle calculations using the projected augmented wave (PAW) formalism of the Kohn-Sham density functional theory [28] within Local density approximation (LDA) [29] and Generalized Gradient approximation (GGA) [30] implanted in Vienna Ab initio Simulation Package (VASP) [31,32] by utilizing the crystallographic model obtained using Neutron diffraction given in supplementary materials [5]. The two calculations yielded equivalent Eigen displacements for the individual modes. The representative displacements of the atoms in the normal phonon modes are shown in fig. 1 along with the Raman spectra collected in different polarization geometry on a commercially obtained single crystal substrate of BaTiO$_3$. Further details of the calculations and characterizations of the single crystal are summarized in the supplementary materials [5]. Experimentally, we have observed E(TO$_1$,TO$_2$,TO$_3$,TO$_4$) modes around 40, 180, 308, 490 cm$^{-1}$, respectively and A(TO$_1$), A(TO$_2$), B$_1$, A(TO$_3$) modes at around 180, 270, 305, 520 cm$^{-1}$, respectively. The mode at the 180 cm$^{-1}$ exhibited asymmetry/dip which was attributed to the interference effect in literature [33]. LO components A(LO$_1$)/E(LO$_1$), E(LO$_2$), A(LO$_2$)/E(LO$_3$), A(LO$_3$)/E(LO$_4$) were observed around 180, 305, 490 and 720 cm$^{-1}$, respectively. The Raman mode positions obtained experimentally and using calculations are tabulated in Table 1 along with the positions reported in previous work on BaTiO$_3$ for comparison.

**Table 1:** Comparison of Experimentally observed, Calculated and Reported Phonon mode position and Character. The 'i' after the numbers for E(TO$_1$) mode indicates that the mode is dynamically unstable in the calculations.

| Current work Experimental | | Current work Calculated Mode frequency | | | Reported Values of Raman mode position | | | | | | |
|---|---|---|---|---|---|---|---|---|---|---|---|
| | | | | | 11 | 14 | 15 | 34 | 35 | 36, 37 | 38 |
| Mode Character | (cm$^{-1}$) | Mode Character | LDA (cm$^{-1}$) | GGA (cm$^{-1}$) | (cm$^{-1}$) | | | | | | |
| E(TO$_1$) | ~40 | E(TO$_1$) | 76.7i | 124.8i | 39 | 161i | 70i | 125i | 128i | | 125i |
| E(TO$_2$) | 180 | E(TO$_2$) | 183.8 | 161.8 | 196 | 167 | 183 | 169 | 182 | 186 | 190 |
| E(TO$_3$) | 305 | E(TO$_3$) | 295.7 | 286.1 | 320 | 284 | 300 | 305 | 316 | 310 | 313 |
| E(TO$_4$) | 490 | E(TO$_4$) | 447.6 | 449.3 | 514 | 457 | 480 | 468 | 487 | | 488 |
| A(TO$_1$) | 180 | A(TO$_1$) | 181.1 | 150.9 | 155 | 161 | 178 | 168 | 168 | 186 | 185 |
| A(TO$_2$) | 270 | A(TO$_2$) | 282.7 | 361.7 | 193 | 302 | 211 | 286 | 373 | 260 | 334 |
| A(TO$_3$) | 520 | A(TO$_3$) | 511.7 | 582 | 554 | 507 | 497 | 517 | 566 | 516 | 543 |
| B$_1$ | 305 | B$_1$ | 295.2 | 270.9 | 282 | 287 | 296 | 300 | 303 | 310 | 311 |
| E(LO$_1$) | 180 | E(LO$_1$) | 179.9 | 157.9 | 182 | 162 | 180 | 169 | 176 | | 184 |
| E(LO$_2$) | 305 | E(LO$_2$) | 295.1 | 283.7 | 308 | 284 | 300 | 305 | 315 | 310 | 312 |
| E(LO$_3$) | 470 | E(LO$_3$) | 425.6 | 428.4 | 462 | 444 | 462 | 454 | 479 | | 473 |
| E(LO$_4$) | 720 | E(LO$_4$) | 533.3 | 658.5 | 699 | 641 | 678 | 687 | 775 | 720 | 724 |
| A(LO$_1$) | 180 | A(LO$_1$) | 192.5 | 184.7 | 193 | 180 | 186 | 174 | 199 | | 203 |
| A(LO$_2$) | 470 | A(LO$_2$) | 451.3 | 443 | 466 | 452 | 465 | 468 | 492 | 477 | 485 |
| A(LO$_3$) | 720 | A(LO$_3$) | 641.3 | 745.6 | 729 | 705 | 697 | 694 | 775 | 725 | 770 |

To establish a direct correlation of calculations with experiments, we performed temperature dependent Raman study (fig. 2).

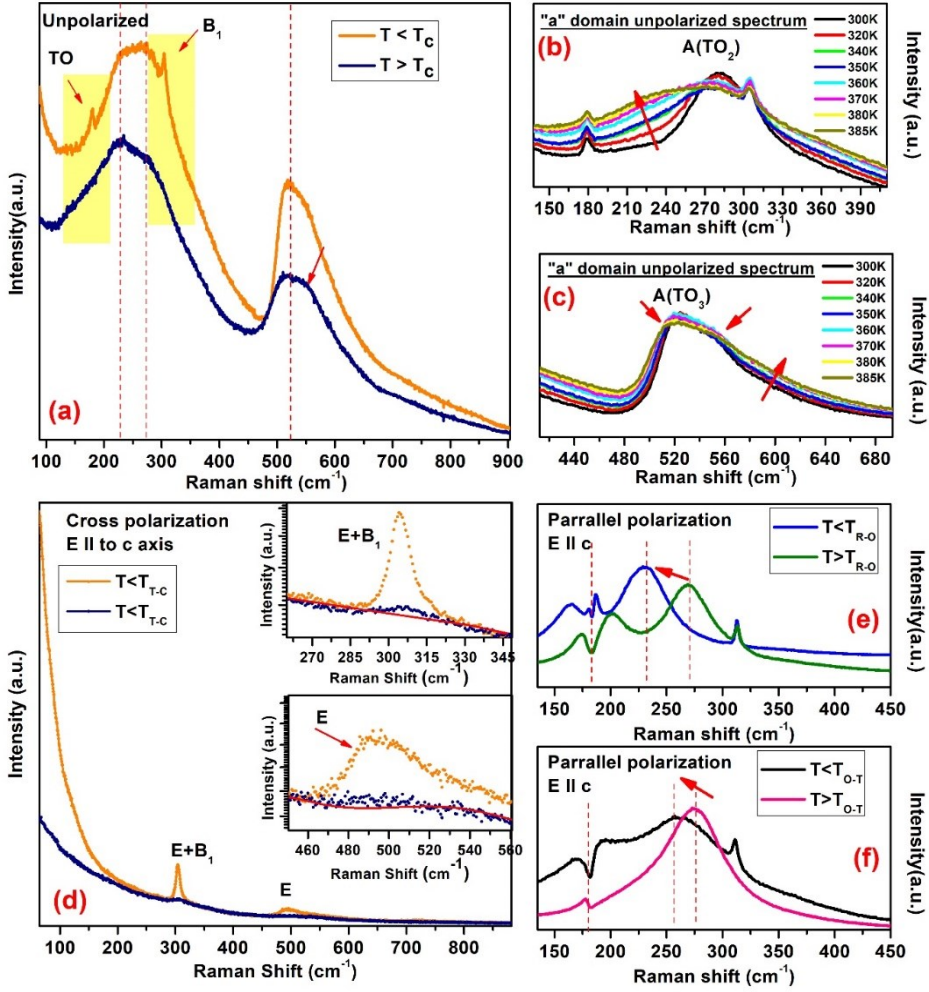

**Fig. 2**: Unpolarized Raman spectra a) across the ferroelectric phase transition $T_c \sim 399K$. Variation in b) $A(TO_2)$ and c) $A(TO_3)$ modes from room temperature to 385 K in an *a*-type domain. d) Raman spectrum in cross polarization geometry across the $T_c$, the insets show magnified view of the $E(TO_3)+B_1+E(LO_2)$ and $E(TO_4)$ modes, respectively. Low temperature Raman spectrum in parallel polarization geometry across e) the rhombohedral to orthorhombic phase transition ($T_{R-O} \sim 197K$) and f) the orthorhombic to tetragonal phase transition ($T_{O-T} \sim 284K$). The absence of $B_1+E(TO_3/LO_2)$ mode, for $T>T_{O-T}$, is due to symmetry constraint.

At the Cubic to tetragonal phase transition, $A(TO_1)$, $E(TO_2)$ and LO modes disappear abruptly while $A(TO_2)$, $A(TO_3)$ mode showed loss in intensity without change in the mode position (fig. 2(a)). For both the $A(TO_2)$ and $A(TO_3)$ modes, we could clearly observe two distinct features. $A(TO_2)$ mode involves a contributions at $\sim 220$ cm$^{-1}$ apart from the well-known feature at $\sim 270$ cm$^{-1}$(fig. 2(a,b)), while $A(TO_3)$ also shows a feature at 560 cm$^{-1}$ apart from well-known feature at $\sim 520$ cm$^{-1}$ (fig. 2(a,c)). Monitoring $A(TO_2)$ mode from room temperature to $T<T_c$, these additional contributions (features) were found to enhance on increasing temperature. Both these features exhibit polarization dependence and shows extinction condition (shown in cross polarization fig. 1 and fig. S2 [5]). The presence of these features in parallel polarization (XX) and its absence in cross (XY) polarization suggests that it has a character identical to A-type normal modes and cannot be attributed to second order contributions. On increasing temperature, both $A(TO_2)$ and $A(TO_3)$ modes showed modulations across $T_B\sim590K$ associated with Burns temperature [39,40]. Interestingly, above 590K, the contribution of the extra feature at 560 cm$^{-1}$ remains significant in $A(TO_3)$ while the contribution of the feature at 520 cm$^{-1}$ diminishes.

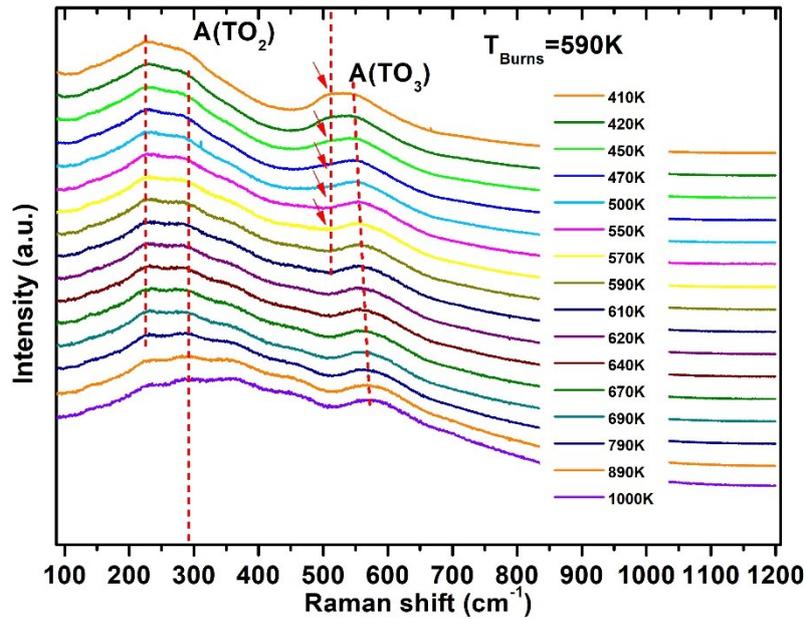

**Fig. 3:** Stacking plot of representative high temperature (410-1000K) Raman spectra collected in unpolarized geometry. Dashed line and arrows are guide to the eyes. Arrow marks the presence of the contribution at 520 cm$^{-1}$.

In order to confirm the Burns temperature, high temperature dielectric measurements were performed from 300-690 K (fig. S5 [5]). However, we could not observe the Burns temperature due to very high losses induced due to enhancement in conductivity at high temperatures (>500K). The deviation in the Curie Weise Law was visible near $T_c$ in dielectric measurements performed in 80-500K temperature interval (fig. S4). The Burns temperature was reported by Second harmonic generation technique which showed deviations [41] in this temperature interval. In order to find out the origin of the extra features observed in our experimental Raman spectroscopy at 220 cm$^{-1}$ and 560 cm$^{-1}$, we studied the Raman spectra of BaTiO$_3$ in different phases. We noted that in rhombohedral phase of the BaTiO$_3$, the $A(TO_2)$ mode occurs at ~220 cm$^{-1}$. Rubio-Marcos *et al* [3] identified rhombohedral domain walls at room temperature through observation of $A(TO_2)$ mode at ~210 cm$^{-1}$ and a feature at 560 cm$^{-1}$. Remarkably, in our experimental study, these additional features which were considered as hallmark of the Ti-dsplacements in the oxygen octahedra along the (111) directions are found well inside the *a*- and *c*-domains in agreement with previous report on monodomain BaTiO$_3$ single crystal [42]. Yu. I. Yuzyuk *et al* [43] argued that these features are associated with local rhombohedral distortion and are remain present in high temperature cubic phase (above T$_c$), while the mode at 520 cm$^{-1}$ associated with tetragonal distortion disappears. Our observation of sudden loss of intensity of most of the Raman modes above $T_C$, (as shown in fig. 2(a), and 3), and suppression of feature at 520 cm$^{-1}$ at Burns temperature while noticeable intensity of the features at 220 cm$^{-1}$ and 560 cm$^{-1}$ are in tune with these two previous reports. To explore it further, we performed in-situ electric field dependent Raman study on the Single crystal BaTiO$_3$. The electric field was applied along (100) direction and the Raman spectrum was collected by mounting crystal in such a way that the incident electric field polarization of the laser is along the (001) direction i.e. parallel to the direction of the dipole moment. Fig. (4) shows Raman spectra collected under the electric field strength of 0 kV/cm and 2.62 kV/cm. On application of the electric field, the polar A(TO$_2$) mode softens (inset of fig. (4)), and the features at ~220 and 560 cm$^{-1}$ near the A(TO$_2$) and A(TO$_3$) modes (marked by arrows and box) show suppression. The modification in the local structure around Ti due to polarization rotation on application of the electric field is reflected in the Raman spectra. The softening of the A(TO$_2$) mode occurs due to the rotation of the dipoles from (001) to (100) direction on application of the electric field along (100). Gradual suppression of 220 and 560 cm$^{-1}$ with increasing electric field occurs due to rotation of (111) directed local off-centered Ti displacement towards (100) direction. This thus confirms presence of off-center displacement of Ti along (111) directions at and above room temperatures.

Our current observation of presence of off-centering of the Ti-atoms at elevated temperatures and the modulations of the Raman modes across Burns temperature (fig. 3) signifies phonon renormalization due to local ordering as reported in ref. [44]. E(TO$_3$/LO$_2$)+B$_1$ at 305 cm$^{-1}$ becomes very broad above phase transition but remain visible till 500K (fig. S4-6) [5]. As for the low frequency region, at the T$_c$, we could observe the drastic fall of the intensity of E(TO$_1$) mode but some broad feature above T$_c$ could still be observed which had been previously related to the relaxation of the Ti atoms [42]. Low temperature Raman spectra showed shift of the A(TO$_2$) mode from 270 to 220 cm$^{-1}$, sharpening of the broad A(TO$_2$) and A(TO$_3$) modes in the rhombohedral phase, and change in the shape of the Raman modes around 180 cm$^{-1}$ due to change in the coupling with A(TO$_2$) mode (fig. 2 (d-f)).

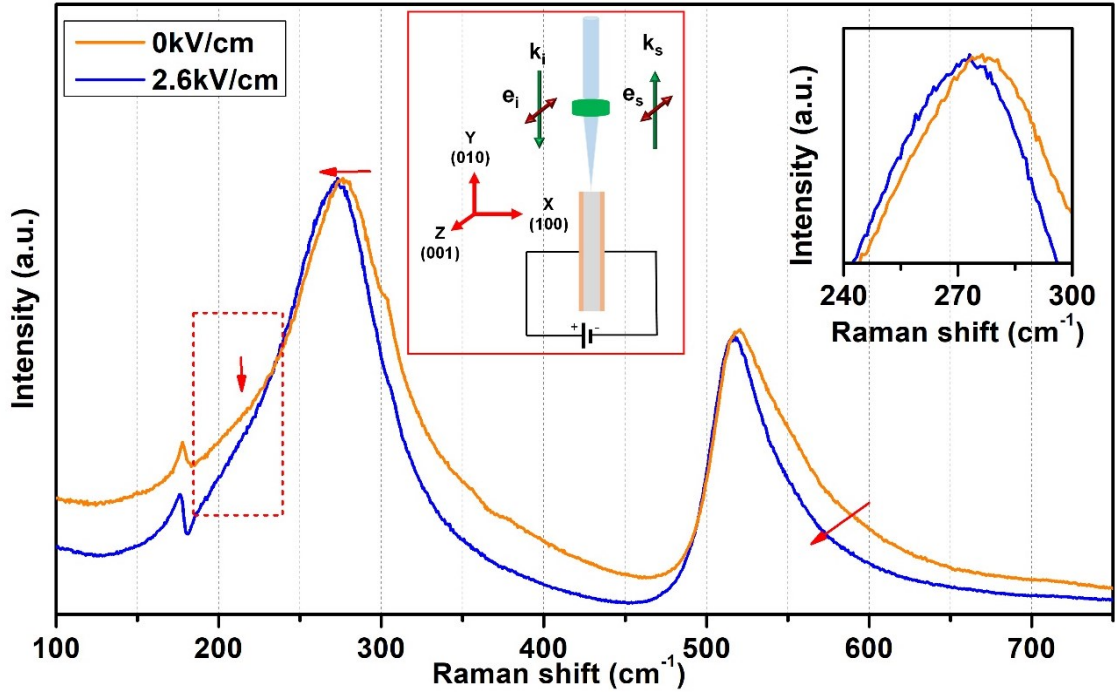

Fig. (4) In-situ electric field dependent Raman spectra at electric field strength of 0, and 2.6 kV on single crystal $BaTiO_3$ with schematic of the experiment in the red box. Inset shows shift in the $A(TO_2)$ mode position. Arrows are guide to the eye and box marks the suppression of the feature at 220 cm$^{-1}$.

## 4. Discussions

The aim of the present work is correct characterization of the polar modes and soft mode. It is well known that the hybridization of Ti-O decides the degree of off-center and consequently the polarization shown by the Titanates. Raman scattering facilitates identifying the off-centering in Titanates through observation of polar modes and hence it is important to correctly characterize them. Further, Raman being a local probe identifies the off-centering even when it is localized in nature unlike in diffraction techniques where the off-centering essentially required to be global to be identified. In order to find the role of Ti-O hybridization and A-site distortion in the induced polar order in $BaTiO_3$, we generated the atomic density maps from the available x-ray and neutron diffraction data [5,25]. To facilitate the study, the atomic density maps of $PbTiO_3$ which possess identical structure but enhanced polarization is also displayed in fig. 5. Atomic density mapping in ferroelectric phase also allows direct identification of the atomic displacements associated with the soft mode as these vibrations are frozen below the ferroelectric transition. Fig. 5 displays the atomic density map in Ti-O plane and A(Ba/Pb)-O plane, respectively. As for the A-O plane, in both the compounds, the motion of the A and O atoms is out-of-phase i.e. the A and O atoms moves opposite to each other. On the other hand, in case of Ti-O plane, it can be seen that the oxygen cage and Ti atom moves opposite to each other i.e. are out-of-phase in $BaTiO_3$ while are in-phase in $PbTiO_3$. This relative motion of the oxygen cage and Ti atom is identified as $E(TO_1)$ normal mode of vibration through our calculations as well as that reported previously [11,14,15,16,17]. From the atomic density maps it is clearly a Slater type ($Ba^+Ti^+O_1^-O_2^-$) vibration in case of $BaTiO_3$ while Last type ($Pb^+Ti^-O_1^-O_2^-$) vibration in case of $PbTiO_3$. As mentioned in introduction, $E(TO_1)$ mode shows characteristics of soft mode in both the compounds. Therefore, we conclude that soft mode in $BaTiO_3$ is Slater type while it is Last type in $PbTiO_3$ in contrary to previous reports where it is termed as Last type mode in both the compounds [24].

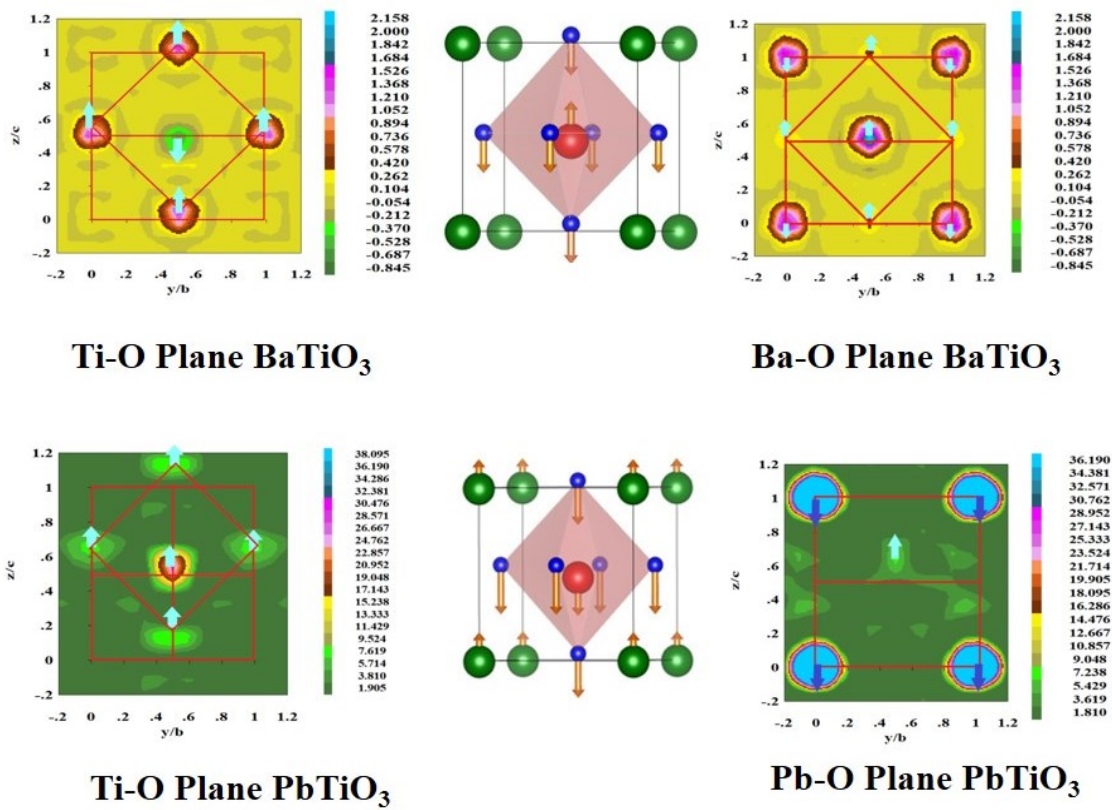

**Fig. 5**: Atomic density maps generated in Ti-O plane and A-O plane projected in YZ plane of the $BaTiO_3$ and $PbTiO_3$, respectively obtained from Neutron and X-ray diffraction data with guided arrows and unit cell imitation.

Now we turn our attention to the identification of the other polar modes. Our atomic density maps and previous reports suggest that in $BaTiO_3$, the off-center displacement of Ti in the oxygen cage is responsible for producing the electric dipoles, while in $PbTiO_3$ the off-center displacement of both the Ti as well as Pb ions in the respective oxygen cages are responsible for the electric dipoles in the ferroelectric phase. According to the old calculations (generally referred in Raman studies [3,11,18, 23]) the $E/A(TO_1)$ modes are defined as Slater type (Ti-O),

E/A(TO$_2$) modes are defined as Last type (Ba-TiO$_3$) while the E/A(TO$_3$) modes are defined as Axe (TiO$_6$) type vibrations. Following these definitions, the absence of the A(TO$_1$) and the presence of the A(TO$_2$) above $T_c$ suggests lattice deformation without Ti displacement (fig. 2(a)). On the contrary, experimental results [45,46,47] provides evidence of presence of off-centered Ti atoms in the oxygen cage above T$_c$. As mentioned above, unlike PbTiO$_3$, BaTiO$_3$ possesses single polar degree of freedom so this leads to a non-physical situation. High pressure x-ray diffused scattering study [48] on BaTiO$_3$ confirmed presence of off-centered Ti atoms locally above 2 GPa, while high pressure Raman scattering study reported presence of only broad A(TO$_2$) and A(TO$_3$) modes above 2 GPa [36,37]. This clearly indicates that A(TO$_2$), A(TO$_3$) modes occurring at 270 and 520 cm$^{-1}$, respectively are related with the relative motion of the Ti in Oxygen octahedra in contrast to previous theoretical studies [11]. Studies on thin films further helps in resolving this discrepancy. It is reported that compressive strain enhances the Ti off-centering in thin films resulting in hardening of the Ti related phonon modes [49]. Raeliarijaona et al [15] calculated the phonon modes under strain and shown hardening of the E(TO$_1$), A(TO$_2$), A(TO$_3$) modes on increasing compressive strain while the E(TO$_2$)/A(TO$_1$) modes remain unchanged. Another clue comes from the electric field dependent Raman studies (fig. 3) i.e. significant change in A(TO$_2$), A(TO$_3$) mode position and intensity indicating direct involvement of the Ti-O bonding. Along with this, our experimental observation of red shift of A(TO$_2$) mode (fig. 2(e-f)) with modifications in local Ti-O bond [46] across low temperature phase transitions establishes a direct correlation of Ti-O bonding with A(TO$_2$). Thus our experimental results suggest that the characterization of the normal modes of vibrations by old calculations do not match with the experimental results. Our calculations suggests that the A(TO$_1$) and E(TO$_2$) modes are Last type while A(TO$_2$) and E(TO$_1$) are Slater type modes in BaTiO$_3$ supporting our experimental observations. Finally, E(TO$_3$)+B$_1$ Raman mode which occurs at 305 cm$^{-1}$ has been associated with ferroelectric phase [23, 50]. In spite of that, these modes remain present above T$_c$ till a very high temperature (T$_c$+100K) as shown in fig. 2, S9-10. Across $T_c$, this mode showed drastic fall in intensity. Our calculations and experimental observations suggest 3 contributions in the Raman feature observed at 305 cm$^{-1}$; E(TO$_3$)/B$_1$/E(LO$_2$). Our calculations show that the Eigen displacements for E(TO$_3$/LO$_2$)+B$_1$ modes are not Axe type as defined in previous literature [11] and the splitting between LO-TO modes is very small. Careful inspection of Eigen displacements reveal out of phase vibrations of the adjacent Basel oxygen atoms in B$_1$ mode compensating off-center displacements of each other and hence is non-polar. While in the case of E(TO$_3$/LO$_2$) modes, the displacements of the only two of the Basel oxygen atoms are out-of-phase relative to the displacements of the apical oxygens and hence these modes are weakly polar. Due to this, only E(TO$_3$/LO$_2$) modes are IR active and show LO-TO character as observed in our calculations. Our calculations suggest that E(TO$_4$)/A(TO$_3$) modes have Axe type atomic displacements as reported for the family of perovskites. Our calculations resulted in Last, Axe and Slater type displacements for E(LO$_1$)/A(LO$_1$) mode at ~180 cm$^{-1}$, E(LO$_3$)/A(LO$_2$) mode around 470 cm$^{-1}$ and E(LO$_4$)/A(LO$_3$) mode around 720 cm$^{-1}$, respectively. Following widely accepted work of Buscaglia and Farhi *et al* [22,51], E(LO$_4$)/A(LO$_3$) modes around 720 cm$^{-1}$ are related to bending and stretching of BO$_6$ octahedra, having mixed A and E character and regarded as indicator of ferroelectricity. Our examination of Eigen displacements for these modes suggest involvement of direct Ti-O ionic displacements i.e. Slater type displacements. Only because of Ti-O involvement, the intensity of this mode modulates across the transition and is not related to the breathing or stretching type vibrations as previously reported in the literature [22,51].

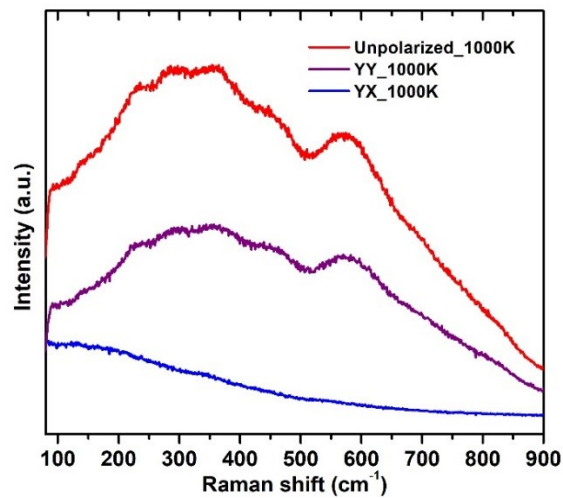

**Fig. 6:** Normalized Raman spectrum of BaTiO$_3$ single crystal at 1000K collected in different polarization geometries.

Presence of Raman modes above T$_c$ represents persisting structural anisotropy in the paraelectric phase [52]. Noticiably, these features remain present even at 1000 K and follows the same polarization selection rules, suggesting first order nature of these features (fig. 6). Previously, these features were identified as arising from second order [4,9,10,42], however, second order contributions are not expected to show any polarization dependence. Anomalously high width of these modes was attributed to dynamical disorder and anharmonicity by Chaves *et al* [4] and Pinczuk *et al* [53], respectively. Direct comparison of Raman spectrum of BaTiO$_3$ with PbTiO$_3$ and K$_{0.5}$Bi$_{0.5}$TiO$_3$, (fig. 7), revealed that even though these systems exhibit large anharmonicity, much larger width is observed for phonons in orthorhombic, tetragonal and cubic phase of BaTiO$_3$ and disordered K$_{0.5}$Bi$_{0.5}$TiO$_3$. This suggests major contribution of dynamical disorder in BaTiO$_3$ as pointed out by Chaves *et al* [4] and cannot be attributed solely to the anharmonicity.

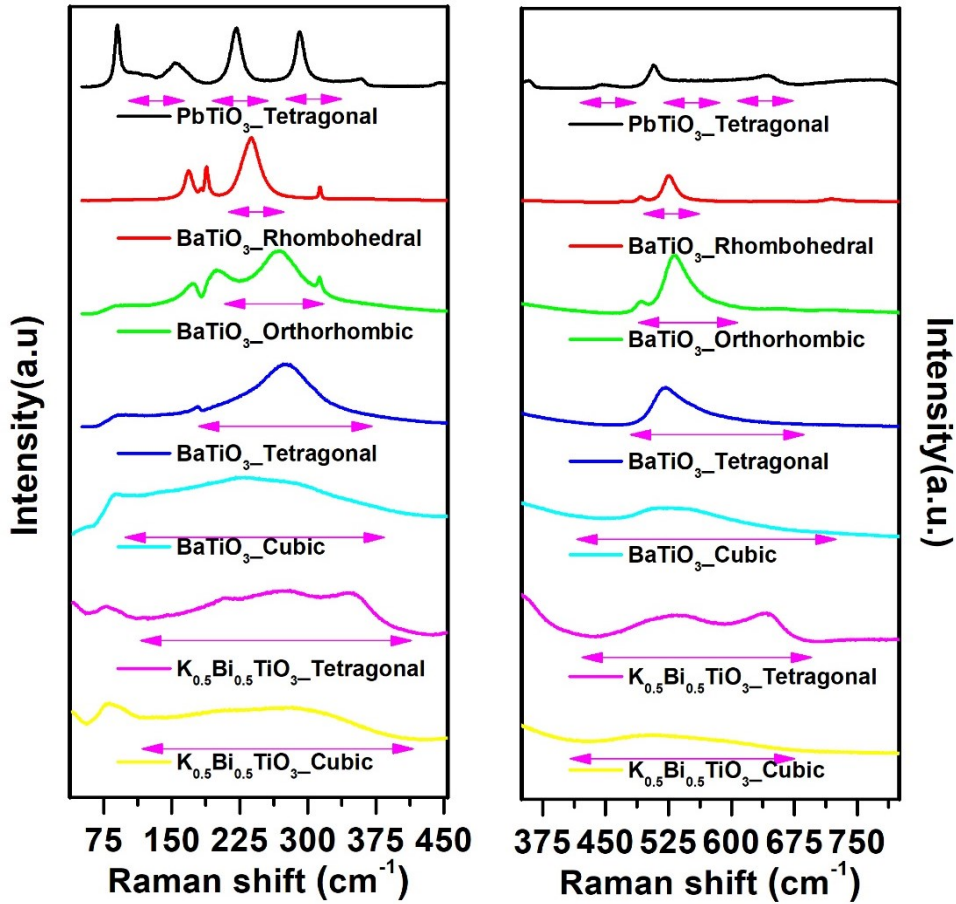

**Fig. 7:** Comparison of width of Raman modes in BaTiO$_3$, PbTiO$_3$ and K$_{0.5}$B$_{0.5}$TiO$_3$ in different phases. Arrows are guide to the eye.

Recent diffused scattering study [54] reported large contribution of anharmonic phonons along with signatures of displacement of the Ti ions along (111) direction which supports this multiple contribution scenario. Further, Hlinka *et al* [19] reported that the disorder component can arise from the A(TO$_2$) mode. Persistence of A(TO$_2$)/A(TO$_3$) modes above T$_c$ (being Ti-O vibrations), verifies that the disorder is related to the dynamics of the Ti atoms. Dominating contributions at ~ 220, 560 cm$^{-1}$ in a Raman spectrum above $T_C$ reflects increased rhombohedral distortion of the unit cell. Modulation of these features under external electric field gives direct evidence of (111) directed displacement of Ti ions as proposed by Comes et al [55]. Our results indicate that disorder component may arise from competition between the on-site potential and intercell (dipolar) interactions as suggested in ref. [56,57]. In Relaxor ferroelectrics, [58,59] high degree of disorder due to different local symmetries induces such deviation and leads to order-disorder dynamics, as exhibited in the present case, due to different local Ti-O hybridization. Coexistence of soft mode and persisting polar modes, absence of shift in Ti-O vibrations above $T_c$ advocates the coexistence of the displacive and order-disorder components [45,46,47]. Though, the Soft mode, is defined in the continuum limit while local modes are defined in quantum limit, so the observations are essentially time and length scale sensitive [60]. The observed nature of transition depends upon the degree of disorder and governed by the superposition of the displacive and disorder components.

## 5. Conclusion

The study provides a new insight in the lattice dynamics of $BaTiO_3$. The polar modes are reclassified by correctly identifying $E(TO_1)$, $A(TO_2)$ as Slater type (Ti-O), $E(TO_2)$, $A(TO_1)$ as Last type ($A-BO_3$) and $E(TO_4)$, $A(TO_3)$ as Axe type ($BO_6$) atomic displacements compatible with the distinct irreducible representations. Multiple contributions in features related to the $A(TO_2)$ and $A(TO_3)$ modes have been recognized as an evidence of off-center displacement of the Ti atoms along the (111) direction in Raman spectrum. Polarized Raman scattering measurements at 1000 K established that all the spectral features seen at high temperatures are due to first order scattering.

Acknowledgement

V.D. acknowledges Dr. S.B. Roy, Sudip Pal, Vinay Kaushik and Pramode Bhakuni for the discussions on the phase transition. V.D. is grateful to Dr. Mukul Gupta, Niti, Seema and Shailesh Kalal for X-ray diffraction. The authors would like to extend special thanks to Dr. Rajeev Ranjan for sharing neutron diffraction data of $BaTiO_3$.